\documentclass[12pt]{article}
\begin{document}
\hfill\vbox{\baselineskip14pt
            \hbox{\bf July 2004-August 2005}}
\baselineskip20pt
\vskip 0.2cm 
\begin{center}
{\Large\bf Clear Experimental Signature of Charge-Orbital density wave in 
Nd$_{1-x}$Ca$_{1+x}$MnO$_{4}$}
\end{center} 
\begin{center}
\large Sher~Alam$^{1}$,~A.T.M.N. Islam$^{2}$,~T.Nagai$^{1}$,~M.~Xu$^{1}$,
Y. Matsui$^{1}$,~and ~I.~Tanaka$^{2}$ 
\end{center}
\begin{center}
$^{1}${\it AML, NIMS, Tsukuba 305-0044, Ibaraki, Japan}\\
$^{2}${\it Crystal Group, Yamanashi University, Yamanashi, Japan}
\end{center}
\begin{center} 
\large Abstract
\end{center}
\begin{center}
\begin{minipage}{16cm}
\baselineskip=18pt
\noindent
  Single Crystals of Nd$_{1-x}$Ca$_{1+x}$MnO$_{4}$ have been prepared by 
the travelling floating-zone method, and possible evidence of a charge
-orbital density wave in this material presented earlier [PRB68,092405 (2003)] 
using High Resolution Electron Microscopy [HRTEM] and Electron 
Diffraction [ED]. In the current note we present direct evidence of 
charge-orbital ordering in this material using heat capacity
measurements. Our heat capacity measurements indicate a clear transition 
consistent with prior observation. We find two main transitions, one at
temperature $T_{_H}=310-314$ K, and other at $T_{_A}=143$ K.
In addition, we may also conclude
that there is a strong electron-phonon coupling in this material.  
\end{minipage}
\end{center}
\vfill
\baselineskip=20pt
\normalsize
\newpage
\setcounter{page}{2}

\section{Introduction}
The transition-metal oxides display a wide variety of interesting
properties. In particular of interest are the 3d transition metals
such as Fe, Cu, Ni, Co and Mn, having a single layered perovskite 
structure, i.e. K$_2$NiF$_4$-type. These materials provide us useful 
insights into the underlying Mott-Hubbard in addition to helping us 
understand the superconducting oxides to which they are similar. 
In a recent note we \cite{alam02} have shown the existence of 
inhomogeneous charge distribution in very good quality
single crystal of LaSrCuO$_{4}$ [LSCO] co-doped with $1\%$ Zn at 
the copper site, using temperature dependent polarized X-ray absorption 
near-edge structure [XANES] spectra. A related single-layered compound 
to the LSCO is LaSrFeO$_{4}$ whose atomic and magnetic structure was 
first elucidated by Souberoux et al.\cite{sou1980},
using X-ray diffraction, Mossbauer spectroscopy and neutron diffraction
in polycrystalline state. Recently Kawanaka \cite{kawa03} have
prepared  a single crystal of  LaSrFeO$_{4}$ by floating-zone
method and demonstrated a spin flop transition in the anti-ferromagnetic
ground state. 

In particular it seems that 3d-electron systems exhibit ordering
and disordering of the fundamental degrees of freedom such as
charge, spin and orbital. In turn it is plausible that this
may be responsible for metal-insulator transition\cite{kawa03}, 
high-T$_c$ superconductivity \cite{alam02, alam02-0}, and colossal 
magneto-resistance \cite{ima98}. A class of materials which show
charge-orbital ordering of $e_g$ electrons are the mixed valent
manganites, the perovskite-type: RE$_{1-x}$AE$_{x}$MnO$_{3}$,
single-layered RE$_{1-x}$AE$_{1+x}$MnO$_{4}$ and double-layered
 RE$_{2-2x}$AE$_{1+2x}$Mn$_{2}$O$_{7}$, where RE=trivalent lanthanides,
and AE=divalent alkaline-earth ions. The real-space ordering pattern,
possibly induced by strong electron-phonon interaction, that is
a cooperative Jahn-Teller effect, can be determined by crystallographic
superstructure \cite{nag03, nag02}. What physical insight can be gained
from these crystallographic superstructure? Here are some examples,
for the case of the over-doped single-layered manganites, the suggested 
models\cite{nag02, lar01} indicate that the Mn valence is not an integer 
but suffers a modulation from site to site. In addition two other 
theoretical proposals of some interest for the over-doped single-layered 
manganites, are the Wigner-crystal\cite{che97} and bi-stripe 
models\cite{mor98}.
The Wigner-crystal model \cite{che97} asserts the stacking of 
Mn$^{3+}$ and Mn$^{4+}$ stripes, such that  Mn$^{3+}$ stripes are 
regularly spaced, in contrast the bi-stripe model \cite{mor98} claims 
the pairing of the Mn$^{3+}$ stripes.

Thus main the purpose of this note is to concentrate on the
experimental results of our heat capacity measurements for the
layered manganites Nd$_{1-x}$Ca$_{1+x}$MnO$_{4}$. Previously two
of us reported \cite{nag03}, using HRTEM and ED presence of 
super-reflections and transverse and sinusoidal modulations
in the charge-orbital ordered phases. The previous measurements \cite{nag03}
were carried out using polycrystalline samples of 
Nd$_{1-x}$Ca$_{1+x}$MnO$_{4}$. Here we report for the first time the
successful growth of single crystals of Nd$_{1-x}$Ca$_{1+x}$MnO$_{4}$
for x=0.67 and x=0.70. The Heat Capacity [HC] was measured using
Quantum Design Physical Property Measurement System [PPMS].

The dc magnetization, was measured using Quantum Design SQUID
magnetometer [MPMS]. This will be reported elsewhere.

This paper is organized as follows, in the 
next section we outline experimental details. This is followed
by section three where the results and discussion of our study
are given. The final section contains the conclusions. 


\section{Experimental}

It should be noted that we have for the first time succeeded in growing
single crystals of Nd$_{1-x}$Ca$_{1+x}$MnO$_{4}$ for x=0.67 and x=0.70.
It is non-trivial to do so, in particular for the case of high calcium
content, i.e. $x \geq 0.72$ it was not possible to obtain single crystals.
 
The crystal growth was carried out using the floating zone technique in a 
four-mirror type infrared image furnace. Feed rod was prepared by solid state
reactions: high purity powders of Nd$_2$O$_3$, CaCO$_3$ and MnO$_2$ were 
mixed in stoichiometric composition and were calcined twice at 
1000 $^{o}$C for 
12 hours with intermediate grinding. In fact about 3\% extra MnO$_2$ was 
mixed to compensate for the vaporisation. The powder was then packed in 
a rubber tube and cold pressed under 300 MPa of pressure to a dense rod 
of ~6 mm in diameter and 50-70 mm in length. The feed rod was then sintered 
at 1500 $^{o}$C for 12 hours. Crystal growth was performed under flowing 
oxygen atmosphere at a speed of 5-10 mm/h.

     The HC of the sample is calculated by subtracting
the addenda measurement from the total heat capacity measurement. The
total HC is the measurement of the HC of the sample, the grease,
and the sample platform. The two measurements-one with and one without
the sample on the sample platform are necessary for accuracy. In order
to ensure the further accuracy of our results, we conducted the
experiment several times, each time repeating the addenda measurement. 
In addition, our several independent measurements are separated
by 20-30 days.We note that automatic subtraction of the addenda, at each 
sample temperature measurement is performed. As we were interested mainly
in the high temperature region, we used H-grease.


\section{Results and Discussion}
Figs.~\ref{fig1}-\ref{fig3} show the results of HC measurements
for the x=0.67 sample. Similarly the results are displayed in
Figs.~\ref{fig4}-\ref{fig6} when the composition is x=0.70. Let 
us call these cases I and II respectively. We
note that both data are self-consistent, after one takes into
account the simple factor due to the mass of the sample. The
mass of the sample in case I was measured to be 5.25 $\pm$ 0.10 mg,
and in case II 8.34 $\pm$ 0.10 mg. The clarity of the measurement
is vividly displayed in Figs.~\ref{fig3} and \ref{fig7}. We can 
identify two main deviations from the normal at temperatures
 $T_{_H}$ and  $T_{_A}$ in Figs.~\ref{fig2} and \ref{fig6}.
The magnitude of the peaks is 5-15 \% above the 'normal' background
in the corresponding temperature range.

       To see the anomalous peaks more clearly, we subtract
the smooth background, and the results are displayed in Figs.~\ref{fig4}
and \ref{fig8}, respectively for the cases I and II. We label
this deviation in Heat Capacity  as $\Delta H$. The location
of the peaks is at $T_{_A}=143.29$ K,  $T_{_H}=314.35$ K for the case I,
and in case II we find $T_{_A}=143.47$ K,  $T_{_H}=310.73$ K. From the
peaks at $T_{_H}$ we find a percentage change, or deviation away from
the background value, of 12.9 \% and 8.3 \% for cases I and II respectively.
The peaks at $T_{_A}$ represent a change of approximately 5.23 \% for case I,
and a 8 \% departure from the background value, for case II.  

       The origin of the strong and distinct anomalous peak in the 
Heat Capacity at T$_{_H}$ is due to charge-orbital ordering, which supports 
our previous work on charge and orbital ordering using electron
microscopy \cite{nag03}. One can also regard this as more direct
evidence for transverse and sinusoidal structural modulations.
It is also tempting to suggest that this feature is related to
a charge-orbital wave of e$_g$ electrons in this material.
The alteration in the manganese valence is nothing but the
variation of the density of e$_g$ electrons. The possible reason
for the modulation of the manganese valence, is the successive change
in the amplitude of the Jahn-Teller distortion in the MnO$_6$ octahedra
with position \cite {nag02, lar01}. It is also possible to interpret 
the modulated structure as an orbital density wave, as mentioned 
previously \cite {nag03}. The density of e$_g$ electrons is constant 
in a pure orbital density wave, however in our case we assume that 
orbital state varies as shown in Fig.~6a of \cite{nag03} between $\pm \pi$.
It is useful to recall the description of orbital state by
the psuedospin space \cite{kan60, kug73}. It is assumed that the 
motion of the psuedospin is confined to the xz plane and orbital state at the 
site i, $\theta_i$ varies according to, 
\begin{eqnarray}
\theta_i &=&\cos(\theta_i/2)|x^2-y^2> +\sin(\theta_i/2)|3z^2-r^2>.\nonumber
\end{eqnarray}
If it is assumed that a pure charge density wave, then the variation
of the orbital state in the direction perpendicular to the stripe
is discrete taking on values  $\pm \pi$. In contrast, the Mn valence
or $e_g$ electron density varies in a sinusoidal manner. Thus it
is more tempting to assume the variation of orbital-density wave
to follow the variation of the charge density wave, as also
proposed in \cite{koi98}. In brief we take the variation of both
charge density and orbital density waves to be sinusoidal.
Incidentally, we can use this interpretation of the sinusoidal
variation of charge-orbital density to also interpret the 
anomaly at T$_{_A}$. We may simply see that in this temperature 
region a further alteration occurs in the charge and magnetic 
state of the material, perhaps involving a further change in 
the local structure, arising from a weak Jahn-Teller effect.
 
It may be remarked that as our simulations of electron 
microscopy \cite{nag03} are consistent with
the Wigner-Crystal model, it is tempting to interpret our HC data
in this context by thinking of it as signalling the formation
of Wigner-Crystal at temperature T$_{_H}$. At the temperature
T$_{_A}$, the anomaly may be seen to arise from the charge-orbital
ordering alteration of the charge and the magnetic state of the 
material. The only work we located for which similar interpretation was given
is \cite{mah03} for the case of cobalt-ate polycrystalline 
Pr$_{0.5}$Co$_{0.5}$ O$_{3}$. It is interesting to note, that two 
anomalies have been reported for the case of the cobalt-ate 
polycrystalline Pr$_{0.5}$Co$_{0.5}$ O$_{3}$,
\cite{mah03}, see Fig. 4 of \cite{mah03}. The anomaly at T$_{_A}$ is 
attributed to the orbital-ordering alteration of the ferromagnetic state.
 
\section{Conclusions}
We have for the first time provided clear experimental
evidence for charge-orbital ordering and related transition using
Heat Capacity measurements in the material 
Nd$_{1-x}$Ca$_{1+x}$MnO$_{4}$. This supports the previous HRTEM
and ED experimental work of two of the authors \cite{nag03}, which gave 
evidence of the presence of a charge-orbital density wave
in layered manganites Nd$_{1-x}$Ca$_{1+x}$MnO$_{4}$.
In addition, the simulations of HRTEM using a transverse rectangle 
structural modulations is consistent with Wigner Crystal model 
\cite{nag03}, and our current results support this viewpoint.
The measurements suggest that the region near and of the anomaly
is  non-perturbative. In addition, we report for the first time the 
growth of single crystals of  Nd$_{1-x}$Ca$_{1+x}$MnO$_{4}$, to our 
knowledge this has not been reported before. 
\section*{Acknowledgments}
The Sher Alam's work is supported by the Japan Society for
for Technology [JST] through the STA fellowship and MONBUSHO
via the JSPS invitation program.

%
\newpage
\begin{figure}
\caption{The Addenda HC [$\mu$ J/K] used for the $x=0.67$
sample.}
\label{fig1}
\end{figure}
\begin{figure}
\caption{The sample HC and the corresponding error for the case
$x=0.67$.}
\label{fig2}
\end{figure}
\begin{figure}
\caption{The main HC anomaly in more detail 
for the case of $x=0.67$.}
\label{fig3}
\end{figure}
\begin{figure}
\caption{The subtracted data, showing the anomalies at
T$_{_H}$  and T$_{_A}$ for the sample with $x=0.67$.}
\label{fig4}
\end{figure}
\begin{figure}
\caption{Results for the HC [$\mu$ J/K] of the Addenda and
the error which was used for the $x=0.70$ sample.}
\label{fig5}
\end{figure}
\begin{figure}
\caption{The sample HC and the corresponding error for the case
when $x=0.70$.}
\label{fig6}
\end{figure}
\begin{figure}
\caption{The main HC anomaly in more detail 
for the case of $x=0.70$.}
\label{fig7}
\end{figure}
\begin{figure}
\caption{The subtracted data, showing the anomalies at
T$_{_H}$  and T$_{_A}$ for the sample with $x=0.70$.}
\label{fig8}
\end{figure}

\end{document}